\documentclass[twocolumn,showpacs,aps,prl,superscriptaddress]{revtex4}

\usepackage{graphicx}
\usepackage{dcolumn}
\usepackage{epsfig}
\usepackage{psfrag}
\usepackage{amsmath}

\newcommand{\BaBarYear}    {09}
\newcommand{\BaBarNumber}  {021}

\newcommand{\SLACPubNumber} {13727}

 \newcommand{\BaBarType}      {PUB}  

\input pubboard/babarsym   

\RequirePackage{xspace}

\newcommand{\jhep}[1]{{\it JHEP}\ #1}

\newcommand{\pvec}{{\bf p}}

\newcommand{\fL}{\ensuremath{f_L}}

\newcommand{\calB}{\ensuremath{{\cal B}}}

\newcommand{\timemsix}{\ensuremath{\times10^{-6}}}

\newcommand{\DE}{\ensuremath{\Delta E}}

\newcommand{\xf}{\ensuremath{{\cal F}}}

\newcommand\etal{{\it et al.}}

\newcommand{\msp}{\ensuremath{\phantom{-}}}

\newcommand{\bfig}{\begin{figure}[htbpc!]}
\newcommand{\efig}{\end{figure}}
\newcommand\bef{\begin{figure}}
\newcommand\edf{\end{figure}}
\newcommand\dbline{\noalign{\vskip 0.10truecm\hrule}\noalign{\vskip 2pt}\noalign{\hrule\vskip 0.10truecm}}

\newcommand\sgline{\noalign{\vskip 0.10truecm\hrule\vskip 0.10truecm}}
\newcommand\beq{\begin{equation}}
\newcommand\eeq{\end{equation}}
\newcommand\bear{\begin{array}}
\newcommand\enar{\end{array}}
\newcommand\beqa{\begin{eqnarray}}
\newcommand\eeqa{\end{eqnarray}}
\newcommand\ben{\begin{enumerate}}
\newcommand\een{\end{enumerate}}

\newcommand{\Kst}{\ensuremath{K^*}}

\newcommand{\Kstp}{\ensuremath{\Kstarp}}

\newcommand{\Kstz}{\ensuremath{\Kstarz}}

   \newcommand{\rhop}{\ensuremath{\rho^+}}
   
   \newcommand{\rhoz}{\ensuremath{\rho^0}}

\newcommand{\bone}{\ensuremath{b_1}}
\newcommand{\bonep}{\ensuremath{b_1^+}}
\newcommand{\bonem}{\ensuremath{b_1^-}}

\newcommand{\bonez}{\ensuremath{b_1^0}}

\newcommand{\fbmrhop}{\ensuremath{\bonem\rhop}\xspace}
\newcommand{\bmrhop}{\ensuremath{\Bz\ra\fbmrhop}\xspace}

\newcommand{\fbprhoz}{\ensuremath{\bonep\rhoz}\xspace}

\newcommand{\fbzrhop}{\ensuremath{\bonez \rhop}\xspace}

\newcommand{\fbzrhoz}{\ensuremath{\bonez \rhoz}\xspace}

\newcommand{\fbpKstz}{\ensuremath{\bonep\Kstz}\xspace}

\newcommand{\fbmKstp}{\ensuremath{\bonem\Kstp}\xspace}

\newcommand{\fbmKstpKppiz}{\ensuremath{\bonem\Kstp_{\Kp\piz}}\xspace}

\newcommand{\fbmKstpKspip}{\ensuremath{\bonem\Kstp_{\KS\pip}}\xspace}

\newcommand{\fbzKstp}{\ensuremath{\bonez\Kstp}\xspace}

\newcommand{\fbzKstpKppiz}{\ensuremath{\bonez\Kstp_{\Kp\piz}}\xspace}

\newcommand{\fbzKstpKspip}{\ensuremath{\bonez\Kstp_{\KS\pip}}\xspace}

\newcommand{\fbzKstz}{\ensuremath{\bonez\Kstz}\xspace}

\newcommand{\rbmrhop}{\ensuremath{-1.8 \pm 0.5 \pm 1.0}}
\newcommand{\sbmrhop}{\ensuremath{-}}
\newcommand{\ulbmrhop}{\ensuremath{1.4}\xspace}

\newcommand{\rbprhoz}{\ensuremath{\msp 1.5\pm 1.5 \pm 2.2}}
\newcommand{\sbprhoz}{\ensuremath{0.4}}
\newcommand{\ulbprhoz}{\ensuremath{5.2}\xspace}

\newcommand{\rbzrhop}{\ensuremath{-3.0\pm 0.9 \pm 1.8}}
\newcommand{\sbzrhop}{\ensuremath{-}}
\newcommand{\ulbzrhop}{\ensuremath{3.3}\xspace}

\newcommand{\rbzrhoz}{\ensuremath{-1.1\pm 1.7^{+1.4}_{-0.9}}}
\newcommand{\sbzrhoz}{\ensuremath{-}}
\newcommand{\ulbzrhoz}{\ensuremath{3.4}\xspace}

\newcommand{\rbmKstpKppiz}{\ensuremath{\msp 1.8\pm 1.9 \pm 1.4}}
\newcommand{\sbmKstpKppiz}{\ensuremath{0.9}}

\newcommand{\rbmKstpKspip}{\ensuremath{\msp 3.2\pm 2.1^{+1.0}_{-1.5}}}
\newcommand{\sbmKstpKspip}{\ensuremath{1.5}}

\newcommand{\rbpKstz}{\ensuremath{\msp 2.9\pm 1.5 \pm 1.5}}
\newcommand{\sbpKstz}{\ensuremath{1.5}}
\newcommand{\ulbpKstz}{\ensuremath{5.9}\xspace}

\newcommand{\rbzKstpKppiz}{\ensuremath{-2.2\pm 3.0^{+5.0}_{-2.3}}}
\newcommand{\sbzKstpKppiz}{\ensuremath{-}}

\newcommand{\rbzKstpKspip}{\ensuremath{\msp 1.6\pm 2.5 \pm 3.3}}
\newcommand{\sbzKstpKspip}{\ensuremath{0.4}}

\newcommand{\rbzKstz}{\ensuremath{\msp 4.8\pm 1.9^{+1.5}_{-2.2}}}
\newcommand{\sbzKstz}{\ensuremath{2.0}}
\newcommand{\ulbzKstz}{\ensuremath{8.0}\xspace}
\newcommand{\UlbzKstz}{\ensuremath{\ulbzKstz\times 10^{-6}}\xspace}

\newcommand{\rbmKstp}{\ensuremath{2.4^{+1.5}_{-1.3} \pm 1.0}}
\newcommand{\sbmKstp}{\ensuremath{1.7}}
\newcommand{\ulbmKstp}{\ensuremath{5.0}\xspace}

\newcommand{\rbzKstp}{\ensuremath{0.4^{+2.0+3.0}_{-1.5-2.6}}}
\newcommand{\sbzKstp}{\ensuremath{0.1}}
\newcommand{\ulbzKstp}{\ensuremath{6.7}\xspace}

\newcommand{\theTitle}{{\boldmath Search for
\B-meson decays to $\bone\rho$ and $\bone\Kst$}} 

\begin{document}

\begin{flushleft}
\babar-\BaBarType-\BaBarYear/\BaBarNumber \\
SLAC-PUB-\SLACPubNumber \\
\end{flushleft}

\title{\theTitle}

%
\author{B.~Aubert}
\author{Y.~Karyotakis}
\author{J.~P.~Lees}
\author{V.~Poireau}
\author{E.~Prencipe}
\author{X.~Prudent}
\author{V.~Tisserand}
\affiliation{Laboratoire d'Annecy-le-Vieux de Physique des Particules (LAPP), Universit\'e de Savoie, CNRS/IN2P3,  F-74941 Annecy-Le-Vieux, France}
\author{J.~Garra~Tico}
\author{E.~Grauges}
\affiliation{Universitat de Barcelona, Facultat de Fisica, Departament ECM, E-08028 Barcelona, Spain }
\author{M.~Martinelli$^{ab}$}
\author{A.~Palano$^{ab}$ }
\author{M.~Pappagallo$^{ab}$ }
\affiliation{INFN Sezione di Bari$^{a}$; Dipartimento di Fisica, Universit\`a di Bari$^{b}$, I-70126 Bari, Italy }
% ...
\collaboration{The \babar\ Collaboration}
\noaffiliation

\date{\today}

\begin{abstract}
We present a search for decays of $B$ mesons to final
states with a \bone\ meson and a $\rho$ or \Kst(892) meson.
The search is based on a data sample consisting of
465 million \BB\ pairs collected by the \babar\ detector at the
SLAC National Accelerator Laboratory. We do not observe any statistically
significant signal. The upper limits we set on the branching 
fractions range from \ulbmrhop\ to \UlbzKstz at the 90\% confidence
level (C.L.), including systematic uncertainties. 
\end{abstract}


\maketitle

Measurements of charmless hadronic \B\ decays are a powerful tool
to test standard model predictions and search for new physics
effects. One of the outstanding problems is represented
by the so called \textit{polarization puzzle} in decays of \B\ mesons
to a pair of spin-one mesons. Simple helicity arguments
predict a longitudinal polarization $f_L$ close to 1. Contrary to this, 
several vector-vector ($VV$) decay modes such as $B\ra \phi\Kst$ \cite{phiKstorig}, 
$B\ra\rhop\Kstz$ \cite{rhopKst0}, and $B\ra\omega\Kst$ \cite{omegaKst} 
exhibit $f_L \sim 0.5$. Possible explanations for 
this puzzle have been proposed within the standard model 
\cite{VVBSMrefs} and in new physics scenarios \cite{nSMetc}.

The measurement of the branching fractions and polarization of
charmless decays of $B$ mesons to an axial-vector and vector meson ($AV$)
may shed light on the size of the amplitudes contributing to charmless
$B$-meson decays and on their helicity structure. 
Theoretical predictions of decay rates have been performed with the
na\"{i}ve factorization (NF) \cite{CMV} and QCD factorization (QCDF) 
\cite{C&Y} approaches.  The NF calculations find
the rates of $B\ra AV$ decays to be smaller than the corresponding
\B\ decays to an axial-vector and pseudo-scalar meson ($AP$).  The
more complete QCDF calculations find the reverse, primarily due to the larger 
decay constants ($\rho$ vs $\pi$ for instance); the expected branching
fractions for the $AV$ modes are substantial in several cases,
as large as $33\times10^{-6}$ for the \bmrhop\ final state.

Additionally, decays of \B\ mesons to charmless $AV$ final states may be sensitive to 
penguin annihilation effects, which tend to enhance certain modes 
while suppressing others. It is thus important to investigate the 
largest possible number of final states.

Measurements of the branching fractions to $AP$ modes $\bone h$, where $h$
denotes a charged or neutral pion or kaon, are presented in Ref. 
\cite{babar_b1P}. The results are in good agreement with the 
predictions of QCDF \cite{C&Y_AP}. 
Searches for the $AV$ decays to the final states $a_1^{\pm}\rho^{\mp}$
and $a_1^+\Kstz$ are presented in Ref. \cite{babar_a1V}, with
upper limits on the branching fractions of $30\times 10^{-6}$ and 
$1.6\times 10^{-6}$ (at the 90\% C.L.), respectively. In this paper 
we search for all charge combinations of decays of a $B$ meson to a 
final state containing a $\bone$ meson and a $\rho$ or $\Kst(892)$ meson. 
No previous searches for these decays have been reported.

The data sample used for these measurements was collected with
the \babar\ detector at the PEP-II asymmetric \epem\ collider 
located at the SLAC National Accelerator Laboratory. The integrated 
luminosity taken at the \FourS\ resonance (center-of-mass energy
$\sqrt{s}=10.58\ \gev$) corresponds to 424 \invfb and is equivalent
to $(465\pm5)\times 10^6$ \BB\ pairs. The \babar\ detector is
described in detail elsewhere \cite{BABARNIM}.

We reconstruct $B$-meson daughter candidates through the decays
$\bone\ra\omega\pi$ (we assume this branching fraction to be 1 
\cite{PDG08}), $\omega\ra\pip\pim\piz$, 
$\rhop\ra\pip\piz$, $\rhoz\ra\pip\pim$, $\Kstz\ra\Kp\pim$, and 
$\Kstp\ra\Kp\piz$ or $\KS\pip$. We impose the following requirements 
on the masses of the selected candidates: $1000 < m(\bone) < 1550\ \mev$, 
$740 < m(\omega) < 820\ \mev$, $470 < m(\rho) < 1070\ \mev$, and 
$755 < m(\Kst) < 1035\ \mev$; these cuts allow some sidebands, 
which help estimating the background level. 
Neutral pions are reconstructed via the decay $\piz\ra\gaga$;
photon candidates with a minimum energy of 50 \mev\ are combined,
and we require the pion energy to exceed 250 \mev in the laboratory
frame. The invariant mass of the $\piz$ candidate is required to be 
in the interval 120$-$150 \mev. 
We select $\KS\ra\pip\pim$ candidates in the mass range $486 < m(\KS) < 510\
\mev$; a kinematic fit constraining the two pion tracks to 
originate from the same vertex is performed and we require 
the $\KS$ flight length to be greater than three 
times its uncertainty.
The daughters of $\bone$, $\omega$, $\rho$ and $\Kst$ are rejected
if their particle identification signatures are consistent 
with those of protons or electrons. $\Kp$ candidates must be 
positively identified as kaons, while $\pip$ must fail kaon 
identification. Unless otherwise stated, charge-conjugate reactions 
are implied.

The helicity angles of the (axial-) vector mesons are measured
in their rest frame. For the $\bone$ candidate, the helicity
angle is defined as the angle between the flight direction of
the pion from the $\bone\ra\omega\pi$ decay and the direction 
of the boost to the $\bone$ rest frame. We define the helicity 
angles of the $\rho$ and $\Kst$ mesons in an analogous manner 
using the direction of the daughter pions [for the $\rho^{\pm}$ 
($\rho^0$) we use the (positively) charged pion]. Finally, the helicity 
angle of the $\omega$ is taken as the angle between the normal to the $3\pi$
decay plane and the direction of the boost to the $\omega$ rest frame. 
To suppress backgrounds originating from low-momentum particles, we
apply the selection criteria summarized in Table \ref{tab:rescuts}. 
Integration over the angle between the $\bone$ and $V$ decay planes 
yields the following expression for the distribution
$F(\theta_A,\theta_V) \propto d^2\Gamma/d\cos\theta_A d\cos\theta_V$  
in the $\bone$ and $\rho/\Kst$ helicity angles $\theta_A$ and $\theta_V$:
\begin{eqnarray} \label{eq:F_ang}
F(\theta_A,\theta_V) &=& \fL\,\left[\cos^2\theta_A+
\left|\frac{C_1}{C_0}\right|^2\sin^2\theta_A\right]\cos^2\theta_V +  \nonumber \\
&&\hspace{-21mm}(1-\fL)\frac{1}{4}\left[\sin^2\theta_A +
\left|\frac{C_1}{C_0}\right|^2(1+\cos^2\theta_A)\right]\sin^2\theta_V.
\end{eqnarray}
Here \fL\ is the longitudinal polarization fraction
$|A_0|^2/\sum{|A_i|^2}$, where $A_i$, $i=-1,0,1$, is a helicity
amplitude of the $B\ra AV$ decay.
The $C_i$ are the helicity amplitudes of
$\bone\ra\omega\pi$; by parity conservation $C_{-1}=C_1$.
The \bone\ decays have been studied in terms of the two parity-allowed
$S$ and $D$ partial wave amplitudes, which have the
measured ratio $D/S = 0.277 \pm 0.027$ \cite{PDG08}.  From this we
obtain the ratio of helicity amplitudes in Eq. \ref{eq:F_ang} \cite{jacobWick} $$
\frac{C_1}{C_0} = \frac{1 + (D/S)/\sqrt{2}}{1 - \sqrt{2}(D/S)}.
$$

\begin{table}[btp]
\begin{center}
\caption{
Selection requirements on the helicity angles of $B$-daughter 
resonances. 
}
\label{tab:rescuts}
\begin{tabular}{lcc}
\dbline
State           & $\rho$/$\Kst$ helicity    &  $\bone$ helicity    \\
\dbline                                        
\fbmrhop        & $-0.50 < \cos\theta_{\rho} < 1.00$ & $-1.0 < \cos\theta_{\bone} < 1.0$ \\ 
\fbzrhop        & $-0.50 < \cos\theta_{\rho} < 0.80$ & $-1.0 < \cos\theta_{\bone} < 0.6$ \\ 
\fbprhoz        & $\msp0.0 < |\cos\theta_{\rho}| < 0.85$ & $-1.0 < \cos\theta_{\bone} < 1.0$ \\ 
\fbzrhoz        & $\msp0.0 < |\cos\theta_{\rho}| < 0.85$ & $-1.0 < \cos\theta_{\bone} < 0.7$ \\ 
\sgline
$\bone^{\pm}\Kst$ &  $-0.85 < \cos\theta_{\Kst} < 1.0$ & $-1.0 < \cos\theta_{\bone} < 1.0$ \\ 
$\bone^{0}\Kst$ &  $-0.85 < \cos\theta_{\Kst} < 1.0$ & $-1.0 < \cos\theta_{\bone} < 0.8$ \\ 
\dbline
\end{tabular}
\vspace{-5mm}
\end{center}
\end{table}
Two kinematic variables characterize the decay of a $B$ meson:
the energy-substituted mass $\mes\equiv\sqrt{s/4-\pvec_B^2}$ and
the energy difference $\DE \equiv E_B-\sqrt{s}/2$, where $(E_B,\pvec_B)$ is
the \B-meson four-momentum vector expressed in the
\FourS\ rest frame. The correlation between the two variables is at the 
few percent level. The resolution on \mes\ is about $2.6\ \mev$, while the 
resolution on \DE\ varies between $20$ and $40\ \mev$ depending on 
the number of $\piz$ mesons in the final state. We select events with 
$5.25 < \mes < 5.29\ \gev$ and $|\DE| < 0.1\ \gev$ except that for \fbzrhop\ we 
require $-0.12 < \DE < 0.10\ \gev$ to allow for the broader signal
distribution when two $\piz$ mesons are present. The average number of $B$ 
candidates per event in the data is between $1.3$ and $1.6$. We choose the candidate with the 
highest value of probability in the fit to the $B$ vertex.

The dominant background originates from
continuum $\epem\ra\qqbar$ events ($q=u,d,s,c$). The angle $\theta_T$
between the thrust axis \cite{thrust} of the $B$ candidate in the \FourS\
rest frame and that of the remaining particles in the event
is a powerful discriminating variable to suppress this 
background. Continuum events peak near 1.0 in the $|\cos\theta_T|$
distribution, while $B$ decays are almost flat. We require
$|\cos\theta_T| < 0.7$ for all the decay modes except
\fbprhoz\, for which we require $|\cos\theta_T| 
< 0.55$, because of substantially higher backgrounds. 
To further reduce continuum background we define a
Fisher discriminant (${\cal F}$) based on five variables related to the event
topology: the polar angles, with respect to the beam axis, of the $B$ candidate
momentum and the $B$ thrust axis; the zeroth and second angular moments
$L_0$ and $L_2$ of the energy flow, excluding the $B$ candidate; and the flavor
tagging category \cite{ccbarK0}.  The first four variables are calculated in the
$\FourS$ rest frame.  The moments are defined by
$L_j = \sum_i p_i\times\left|\cos\theta_i\right|^j,$ where $\theta_i$ 
is the angle with respect to the $B$ thrust axis of track or neutral 
cluster $i$, $p_i$ is its momentum. The Fisher variable  provides 
about one standard deviation of separation between $B$-decay events 
and combinatorial background.

The signal yields are obtained from extended maximum likelihood
fits to the distribution of the data in nine observables: \DE, \mes,
\xf, $m_k$, and $\cos\theta_k$; 
$m_k$ and $\theta_k$ are the mass and the helicity angle of meson $k$
($k = \bone,\; \omega$, and either $\rho$ or $\Kst$). For  
each category $j$ (signal, \qqbar\ background and 
backgrounds originating from \BB\ decays), we define the probability
density functions (PDFs) ${\cal P}_j(x)$ for the variable $x$,
with the associated likelihood $\cal L$:
\begin{eqnarray}
&&{\cal L} = \frac{e^{-(\sum_j Y_j)}}{N!} \prod_{i=1}^N \sum_j Y_j \times \label{eq:totalL}\\
&&{\cal P}_j(\DE^i){\cal P}_j(\mes^i) {\cal P}_j(\xf^i)
\prod_k \left( {\cal P}_j(m_{k}^i){\cal P}_j(\cos\theta_{k}^i) \right),
\nonumber
\end{eqnarray}
where $Y_j$ is the event yield for component $j$ and $N$ is the
number of events entering the fit. We separately model correctly
reconstructed signal events and self-crossfeed (SXF) events, which
are signal events for which particles are incorrectly assigned to
the intermediate resonances, or particles from the rest of the 
event are selected. The fraction of SXF is 0.33-0.57 depending 
on the final state. The signal yields for the branching fraction 
measurements are extracted with the use of correctly reconstructed 
signal events only.

Backgrounds originating from $B$ decays are modeled from Monte
Carlo (MC) simulation \cite{geant}. We select the most significant
charmless modes (20-40 for each signal final state) entering 
our selection and build a sample taking into account measured
branching fractions or theoretical predictions. The expected
charmless \BB\ background yield varies between 26 and 330 events, depending
on the final state. The samples include the nonresonant 
contributions affecting $\bone\rho$ ($\bone\Kst$), measured 
in our data by fitting the central regions of the $\bone \pi \pi$ 
($\bone K \pi$) and $\omega \pi \rho$ ($\omega \pi \Kst$) Dalitz
plots. We assume the probability of the four-body nonresonant
contributions to pass our selections to be negligible.
We do not introduce a component modeling $B$ decays to charmed 
mesons, since this background is effectively absorbed in the
\qqbar\ component.

For the $\Kst$ modes we consider the potential background contribution
originating from $K\pi$ $S$-wave entering the $\Kst(892)$
selection. We model this component using the LASS model 
\cite{LASS,Latham} which accounts for the interference between
the $\Kst_0(1430)$ resonance and the nonresonant component.
The shape of the $K\pi$ invariant mass is kept fixed to the
results found in \cite{LASS}; we fit for the LASS yield
in the range $1035 < m(K\pi) < 1550\ \mev$ and extrapolate
the expected yield to the signal region $755 < m(K\pi) < 1035\ 
\mev$. We find yields that are consistent with zero,
ranging from -56 to 65 events. We fix this yield to zero if it is
negative and take the estimated value otherwise.

PDF shapes for signal, $K\pi$, and \BB\ backgrounds are 
determined from fits to MC samples, while for the \qqbar\
background we use data samples from which the signal region,
$5.27 < \mes\ < 5.29\ \gev$ and $|\DE| < 0.075\ \gev$, is
excluded. The calibration of $\mes$ and $\DE$ is
validated using high-statistics data control samples
of $B$ decays to charmed mesons with similar topologies
(e.g. $B\ra D(K\pi\pi)\pi$, $B\ra D(K\pi\pi)\rho$).

We use linear combinations of polynomial, exponential, and 
Gaussian functions to parameterize most of the PDFs. For 
\qqbar\ background, we adopt a parameterization motivated
by phase-space arguments \cite{Argus}. 

We allow the most important parameters of the \qqbar\ background
to vary in the fit, along with the signal yield. Given that the
signal yields we extract are small, we cannot vary the
longitudinal polarization fraction $f_L$. Since no
strong theoretical predictions exist about its value, we impose 
$f_L = 0.5$ and vary it within the physical range to evaluate the 
systematic uncertainty. We do not include the SXF component in fits with signal 
yields that are consistent with zero to avoid instabilities 
in the SXF fitted yield. In the case of the \fbzKstz mode, where the 
(statistical only) signal significance exceeds three standard deviations,
we retain the SXF component, fixing its yield to correspond to the rate
given by the simulation for its size compared with the signal yield. 
In this case, introducing the SXF component causes the signal 
yield to vary by a small fraction of the statistical error.

\begin{figure}[!b]
\includegraphics[angle=0,width=0.49\textwidth]{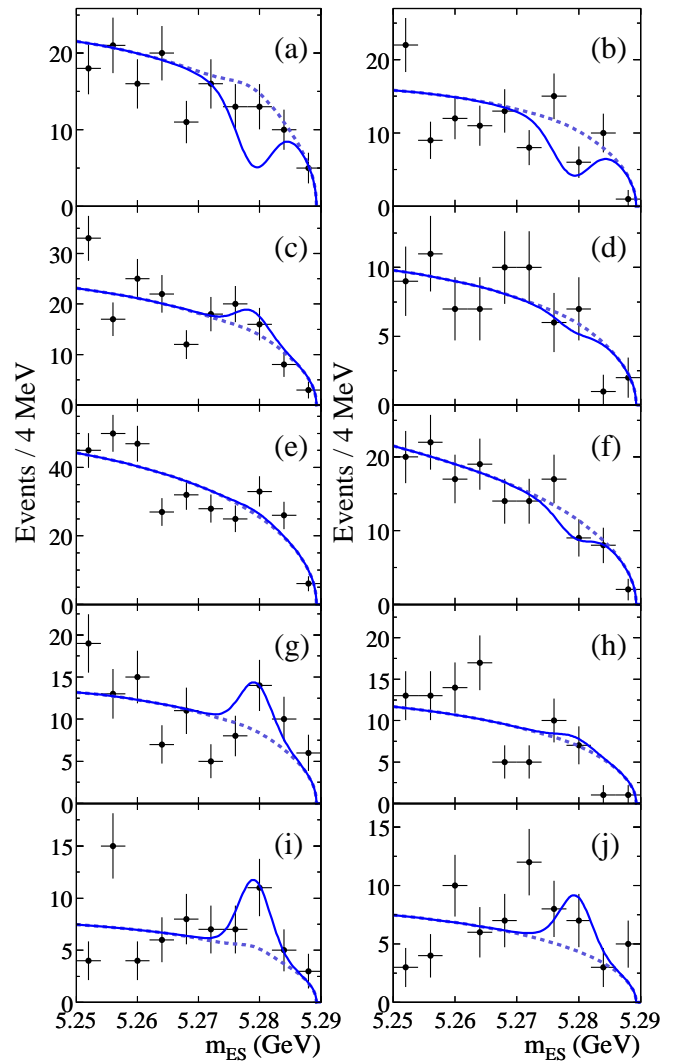}\\
 \caption{\label{fig:proj_mes}Projections onto \mes\ for the modes (a) 
\fbmrhop, (b) \fbzrhop, (c) \fbprhoz, (d) \fbzrhoz, (e) \fbmKstpKppiz,
(f) \fbzKstpKppiz, (g) \fbmKstpKspip, (h) \fbzKstpKspip,
(i) \fbpKstz, (j) \fbzKstz. Points with error bars represent the data, 
the solid (dashed) line represents the total (sum of the backgrounds) 
fitting function. The background is suppressed by a cut on $\ln{\cal L}$, 
optimized separately for each final state.}
\end{figure}

To evaluate the potential bias $Y_0$ that arises from neglecting 
the correlations among the variables entering the fit, we perform 
fits to ensembles of simulated experiments. Each such experiment has 
the same number of signal and background events as the data; \qqbar\
events are generated from the PDFs, while for the other categories
events are taken from fully simulated MC samples.

\begin{table*}[btp]
\begin{center}
\caption{Signal yield $Y$ and its statistical uncertainty, bias $Y_0$,
detection efficiency $\varepsilon$, significance $S$ (including 
systematic uncertainties) and central value of the branching
fraction $\cal B$ with associated upper limit (U.L.) at 90\% C.L.
The efficiency $\varepsilon$ takes into account the product of the
branching fractions of the intermediate resonances.}
\label{tab:results}
\begin{tabular}{lcccccc}
\dbline
Mode            & $Y$      &    $Y_0$   &$\varepsilon~~~$ & $S~$       &  \calB\ & \calB\ U.L. \\
                & (evts) & (evts)   &(\%)$~$       &$(\sigma)$  & $(10^{-6})$  &  $(10^{-6})$ \\
\sgline
\fbmrhop        & $-33\pm10$ & $\msp4\pm2$   & 3.0 & \sbmrhop  & \rbmrhop\ & \ulbmrhop \\
\fbzrhop        & $-18\pm5\:\:$ & $-4\pm2$ & 1.1 & \sbzrhop  & \rbzrhop\ & \ulbzrhop \\
\fbprhoz        & $\msp37\pm25$ & $\msp8\pm4$ & 3.6 & \sbprhoz  & \rbprhoz\ & \ulbprhoz \\
\fbzrhoz        & $\:\:-8\pm19$ & $\msp5\pm3$   & 2.4 & \sbzrhoz  & \rbzrhoz\ $\:\:$ & \ulbzrhoz \\
\sgline
\fbmKstp        &        &    &   & \sbmKstp  & \rbmKstp\ & \ulbmKstp \\
~~\fbmKstpKppiz   & $\msp3\pm8$ & $-5\pm3$ & 0.8 & \sbmKstpKppiz  & \rbmKstpKppiz & \\
~~\fbmKstpKspip   & $\:17\pm9$  & $\msp4\pm2$ & 0.9 & \sbmKstpKspip\ & \rbmKstpKspip\ $\:\:$ & \\
\fbzKstp        &        &    &   & \sbzKstp  & \rbzKstp\ $\:\:\,$ & \ulbzKstp \\
~~\fbzKstpKppiz   & $-8\pm7$   & $-3\pm2$  & 0.5 & \sbzKstpKppiz  & \rbzKstpKppiz $\:\:\:\,$ & \\
~~\fbzKstpKspip   & $\msp3\pm4$ & $\msp0\pm0$ & 0.4 & \sbzKstpKspip  & \rbzKstpKspip & \\
\fbpKstz        & $\:\:\:55\pm21$   & $\:15\pm8$ & 2.8 & \sbpKstz  & \rbpKstz\ & \ulbpKstz \\
\fbzKstz        & $\:\:\:30\pm15$   & $-6\pm3$ & 1.7 & \sbzKstz  & \rbzKstz\ $\:\:$ & \ulbzKstz \\
\dbline
\end{tabular}
\vspace{-5mm}
\end{center}
\end{table*}

We compute the branching fraction $\cal{B}$ for each mode
by subtracting $Y_0$ from the fitted signal yield $Y$ and 
dividing by the efficiency $\varepsilon$ and the number of
\B\ mesons in our data sample. We assume the branching fractions of the \FourS\ to
\BpBm\ and \BzBzb\ to be each 50\%, consistent with
measurements \cite{PDG08}. We evaluate $\varepsilon$ from 
signal MC samples, taking into account the difference
in reconstruction efficiency for longitudinally and transversely
polarized events. For the $\Kstp$ modes, we combine the branching
fraction results from the two sub-modes by adding their $-2\ln{\cal L}$ curves. 
The significance $S$ is computed from the difference between the value
of $-2\ln{\cal L}$ at zero signal and its minimum value. The results 
are summarized in Table \ref{tab:results} while in Fig. \ref{fig:proj_mes} 
we show the projection plots onto the \mes\ variable for the ten final 
states we investigated. We do not observe a statistically significant 
signal for any of the eight decay modes. We quote 
upper limits on their branching fractions at the 90\% C.L., taken 
as the branching fractions below which lie 90\% of 
the totals of the likelihood integrals, constraining the branching 
fractions to be positive. The systematic uncertainties are taken 
into account by convolving the likelihood function with a Gaussian 
of width corresponding to the total systematic uncertainties.

We study the systematic uncertainties due to imperfect
modeling of the signal PDFs by varying the relevant parameters
by their uncertainties, derived from the consistency of fits to 
data and control samples (the systematic uncertainty on the signal yield
varies from 0.6 to 4.1 events, depending on the final state). 
The uncertainty due to the bias correction 
is taken as the sum in quadrature of half the correction itself and 
its statistical uncertainty (0.4-7.5 events). We vary
the yield of the \BB\ backgrounds by $\pm 50\%$ (the resulting
uncertainty is 0.1-8.5 events) and the yield of the $S$-wave $K\pi$ 
component by the larger of $\pm 100\%$ of the extrapolated yield 
and its statistical uncertainty (0.2-14.3 events). The asymmetric 
uncertainty associated with \fL\ is estimated by taking the 
difference in the measured ${\cal B}$ between the nominal fit 
($\fL = 0.5$) and the maximum and minimum values found in the 
scan along the range $[0,\,1]$. We divide these values 
by $\sqrt{3}$, motivated by our assumption of a flat prior for \fL\ in its 
physical range; this is one of the largest sources of systematic
uncertainty, ranging from 0.1 to 3.6 \timemsix. 
Another large source of uncertainty is imperfect knowledge of the SXF 
fraction; based on studies of control samples performed in similar
analyses, we assign a 5\% multiplicative systematic uncertainty 
on the SXF fraction (relative to correctly reconstructed signal) 
for each $\piz$ in the final state. Other uncertainties arise 
from the reconstruction of charged particles (0.4\% per track), 
$\KS$ (1.5\%), and $\piz$ mesons (3\% for $\piz$); the uncertainty 
in the number of $B$ mesons is 1.1\%. 

In summary, we present a search for decays of $B$ mesons to $\bone\rho$
and $\bone\Kst$ final states. We find no significant signals and determine upper
limits at 90\% C.L. between \ulbmrhop\ and \UlbzKstz, including
systematic uncertainties. Though these results are in agreement with the
small predictions from na\"{i}ve factorization calculations \cite{CMV}, 
they are much smaller than the predictions from the more complete QCD 
factorization calculations \cite{C&Y}.  The fact that the branching fractions 
for these $AV$ modes are smaller than our previously measured $AP$ modes 
\cite{babar_b1P} is surprising given that the opposite is expected 
based on the ratio of the vector and pseudoscalar decay constants.

\par We are grateful for the excellent luminosity and machine conditions
provided by our \pep2\ colleagues, 
and for the substantial dedicated effort from
the computing organizations that support \babar.
The collaborating institutions wish to thank 
SLAC for its support and kind hospitality. 
This work is supported by
DOE
and NSF (USA),
NSERC (Canada),
CEA and
CNRS-IN2P3
(France),
BMBF and DFG
(Germany),
INFN (Italy),
FOM (The Netherlands),
NFR (Norway),
MES (Russia),
MEC (Spain), and
STFC (United Kingdom). 
Individuals have received support from the
Marie Curie EIF (European Union) and
the A.~P.~Sloan Foundation.


\renewcommand{\baselinestretch}{1}


\begin{thebibliography}{99}

\bibitem{phiKstorig}
\babar\ Collaboration, B. Aubert \etal, \jprl{91}, 171802 (2003);
Belle Collaboration, K.F.~Chen \etal, \jprl{91}, 201801 (2003).

\bibitem{rhopKst0}
Belle Collaboration, K.~Abe \etal, \jprl{95}, 141801 (2005);
\babar\ Collaboration, B. Aubert \etal, \jprl{97}, 201801 (2006).

\bibitem{omegaKst}
\babar\ Collaboration, B. Aubert \etal, \jprd{97}, 052005 (2009).

\bibitem{VVBSMrefs}
C.W.~Bauer \etal, \jprd{70}, 054015 (2004);
P.~Colangelo, F.~De Fazio, and T.N.~Pham, \plb{597}, 291 (2004);
A.L.~Kagan, \plb{601}, 151 (2004);
M.~Ladisa \etal, \jprd{70}, 114025 (2004);
H.~Y.~Cheng, C.~K.~Chua, and A.~Soni, \jprd{71}, 014030 (2005);
H.-n.~Li and S.~Mishima, \jprd{71}, 054025 (2005);
H.-n.~Li, \plb{622}, 63 (2005).

\bibitem{nSMetc}
A.~K.~Giri and R.~Mohanta, \jprd{69}, 014008 (2004); 
E.~Alvarez \etal, \jprd{70}, 115014 (2004); 
P.~K.~Das and K.~C.~Yang, \jprd{71}, 094002 (2005);
C.-H.~Chen and C.-Q.~Geng, \jprd{71}, 115004 (2005); 
Y.-D.~Yang, R.~M.~Wang and G.~R.~Lu, \jprd{72}, 015009 (2005);
A.~K.~Giri and R.~Mohanta, \epjc{44}, 249 (2005);
S.~Baek \etal, \jprd{72}, 094008 (2005);
W.~Zou and Z. Xiao, \jprd{72}, 094026 (2005);
Q.~Chang, X.-Q.~Li, and Y.~D.~Yang, \jhep{0706}, 038 (2007).

\bibitem{CMV}
G.~Calderon, J.~H.~Munoz, and C.~E.~Vera, \jprd{76}, 094019 (2007).

\bibitem{C&Y}
H.-Y.~Cheng and K.-C.~Yang, \jprd{78}, 094001 (2008).

\bibitem{babar_b1P}
\babar\ Collaboration: B.\ Aubert \etal, \jprl{99}, 241803 (2007); 
\jprd{78}, 011104(R) (2008). 

\bibitem{C&Y_AP}
H.-Y.~Cheng and K.-C.~Yang, \jprd{76}, 114020 (2007).

\bibitem{babar_a1V}
\babar\ Collaboration: B.\ Aubert \etal, \jprd{74}, 031104 (2006);
arXiv:0808.0579 [hep-ex] (2008).

\bibitem{PDG08}
C.~Amsler \etal, \plb{667}, 1 (2008). 

\bibitem{BABARNIM}
\babar\ Collaboration: B.\ Aubert \etal, \nima{479}, 1 (2002).

\bibitem{jacobWick}
M. Jacob and G. C. Wick, \annp{281}, 774 (2000) (originally {\it
op. cit.}, {\bf 7}, 404 (1959)).

\bibitem{thrust}
S. Brandt \etal, \jprl{12}, 57 (1964); E. Farhi, \jprl{39}, 1587 (1977).

\bibitem{ccbarK0}
\babar\ Collaboration, B.\ Aubert \etal, \jprl{99}, 171803 (2007).

\bibitem{geant}
The \babar\ detector Monte Carlo simulation is based on GEANT4
[S. Agostinelli \etal, \nima{506}, 250 (2003)] and EvtGen [D.~J.~Lange,
\nima{462}, 152 (2001)].

\bibitem{LASS}
LASS Collaboration, D.~Aston \etal, \npb{296}, 493 (1988).

\bibitem{Latham}
\babar\ Collaboration, B.\ Aubert \etal, \jprd{72}, 072003 (2005); 
{\bf 74}, 099903(E) (2006).

\bibitem{Argus}
ARGUS Collaboration, H.~Albrecht \etal, \plb{241}, 278 (1990).


\end{thebibliography}
\end{document}